\documentclass[twocolumn,preprintnumbers,amsmath,amssymb,superscriptaddress,subeqn,prl]{revtex4-1}
\usepackage{amsmath}
\usepackage{stmaryrd}
\usepackage{graphicx}
\usepackage{feynmf}
\usepackage{times}
\usepackage{color} 
\usepackage[colorinlistoftodos]{todonotes}
\usepackage{titlesec}
\newcommand*{\justifyheading}{\raggedright}
\titleformat{\section}
  {\normalfont\Large\mdseries\justifyheading}{\thesection}{1em}{}
\titleformat{\subsection}
  {\normalfont\large\mdseries\justifyheading}{\thesubsection}{1em}{}
\titleformat{\subsubsection}
  {\normalfont\normalsize\mdseries\justifyheading}{\thesubsubsection}{1em}{}
  \renewcommand\thesection{\arabic{section}}
\renewcommand\thesubsection{\thesection.\arabic{subsection}}
\renewcommand\thesubsubsection{\thesubsection.\arabic{subsection}}

\makeatletter
\renewcommand*{\@fnsymbol}[1]{\ensuremath{\ifcase#1\or \dagger\or *\or  \ddagger\or
\mathsection\or \mathparagraph\or \|\or **\or \dagger\dagger
\or \ddagger\ddagger \else\@ctrerr\fi}}
\makeatother

\begin{document}

\title{Defect-induced modification of low-lying excitons and valley selectivity \\in monolayer transition metal dichalcogenides}

\author{Sivan Refaely-Abramson}
\thanks{These two authors contributed equally.}
\affiliation{Department of Physics, University of California at Berkeley, Berkeley, CA 94720 , USA}
\affiliation{Molecular Foundry, Lawrence Berkeley National Laboratory, Berkeley, CA 94720, USA}

\author{Diana Y. Qiu}
\thanks{These two authors contributed equally.}
\affiliation{Department of Physics, University of California at Berkeley, Berkeley, CA 94720 , USA}
\affiliation{Materials Sciences Division, Lawrence Berkeley National Laboratory, Berkeley, CA 94720, USA}



\author{Steven G. Louie}
\thanks{Corresponding authors: sglouie@berkeley.edu and jbneaton@lbl.gov}
\affiliation{Department of Physics, University of California at Berkeley, Berkeley, CA 94720 , USA}
\affiliation{Materials Sciences Division, Lawrence Berkeley National Laboratory, Berkeley, CA 94720, USA}

\author{Jeffrey B. Neaton}
\thanks{Corresponding authors: sglouie@berkeley.edu and jbneaton@lbl.gov}
\affiliation{Department of Physics, University of California at Berkeley, Berkeley, CA 94720 , USA}
\affiliation{Molecular Foundry, Lawrence Berkeley National Laboratory, Berkeley, CA 94720, USA}
\affiliation{Kavli Energy Nanosciences Institute at Berkeley, Berkeley, CA 94720, USA}

\date{\today}
\begin{abstract}
We study the effect of point-defect chalcogen vacancies on the optical properties of monolayer transition metal dichalcogenides using \textit{ab initio} GW and Bethe-Salpeter equation calculations. We find that chalcogen vacancies introduce unoccupied in-gap states and occupied resonant defect states within the quasiparticle continuum of the valence band. These defect states give rise to a number of strongly-bound defect excitons and hybridize with excitons of the pristine system, reducing the valley-selective circular dichroism. Our results suggest a pathway to tune spin-valley polarization and other optical properties through defect engineering.
\end{abstract}
\maketitle

\begin{figure*}
\begin{center}
\includegraphics[width=0.95\textwidth]{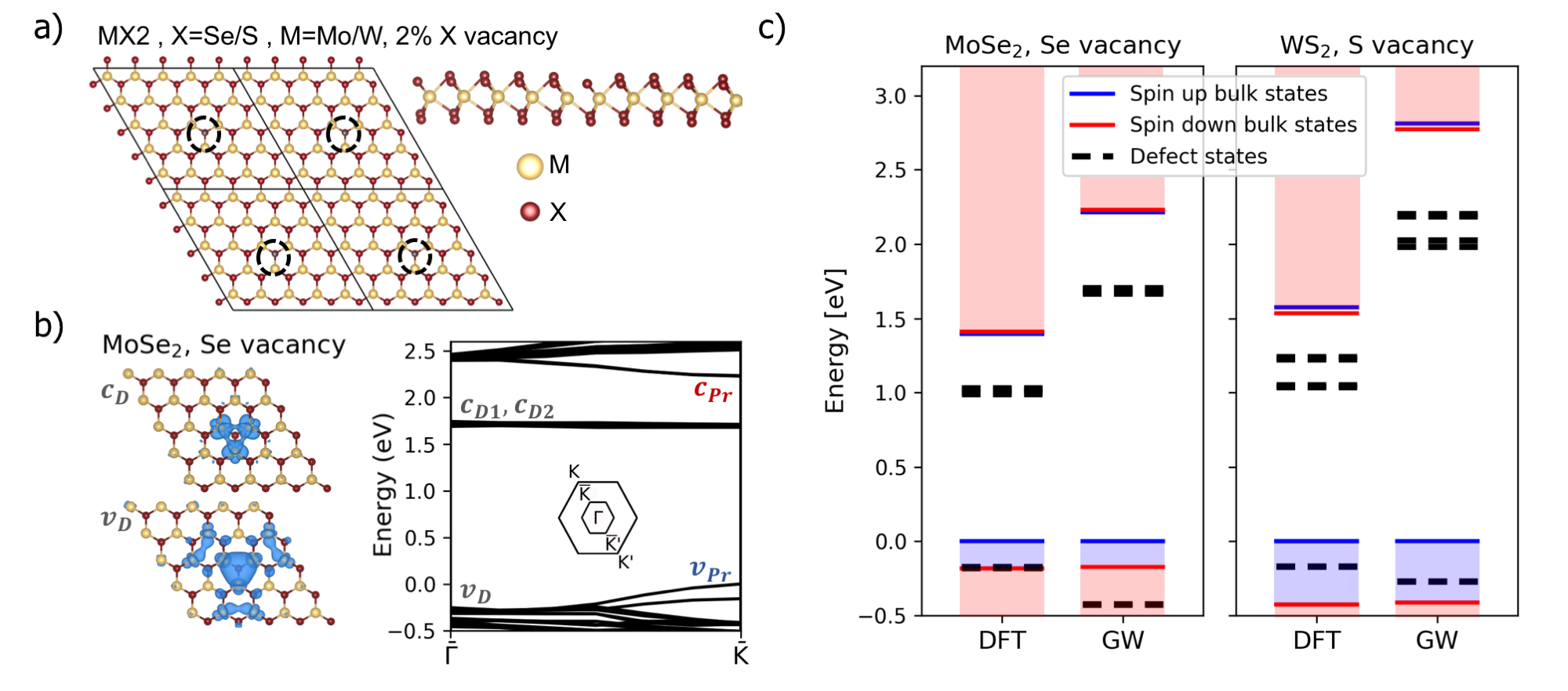}
\centering
\caption{(a) Top and side view of 5$\times$5 TMD supercells with one chalcogen vacancy. Vacancy sites are circled by a black dotted line. (b) left: isosurfaces of the occupied (bottom) and unoccupied (top) wavefunctions associated with the defect states in a 5$\times$5 supercell of MoSe$_{2}$. Right: The quasiparticle bandstructure of MoSe$_2$, along the $\bar{\Gamma}$ to $\bar{K}$ line in the supercell BZ (inset), calculated using G$_0$W$_0$@LDA, including SOC corrections. Defect/pristine-like states are labeled as $v_{D/Pr}$ ($c_{D/Pr}$) for occupied (unoccupied) states. The inset shows a schematic (not to scale) of the supercell BZ compared to the BZ of a single unit cell of a pristine TMD. (c) Defect-state energy levels, calculated within DFT (LDA) and G$_0$W$_0$ for MoSe$_2$ (left) and WS$_2$ (right). Defect states are shown by black dashed lines, and the bulk states are shown in red and blue shaded regions for the spin up and spin down spin-orbit split bands.}
\label{fig:fig1}
\end{center}
\end{figure*}

Monolayer transition metal dichalcogenides (TMDs) are the subject of intense interest due to their remarkable electronic and optical properties, including a direct gap in the visible~\cite{Mak2010,Splendiani2010} and strongly bound excitons and trions \cite{Ashwin2012, Qiu2013, Mak2013, Chernikov2014, Ugeda2014, Klots2014, Zhu2014, Ye2014,Hill2015,Qiu2016} that exhibit novel physics, such as linear dispersion with topological character~\cite{Qiu2015,Wu2015,Yu2014}. Monolayer TMDs also feature locking of valley and spin degrees of freedom, leading to selective excitation of states in different valleys by left- and right-hand circularly polarized light \cite{Cao2012,Xiao2012,Mak2012,Zeng2012}. In particular, the low-energy optical spectra of few-layer TMDs are dominated by two sharp excitonic peaks, referred to as ''A'' and ''B''. These peaks are spin-orbit split excitations, with each peak corresponding to two degenerate excitons arising respectively from transitions between the highest valence bands and lowest conduction bands of the same spin type in the K and K' valleys of the Brillouin zone (BZ).  These excitons exhibit strong valley polarization: upon excitation with circularly polarized light, the emitted light from the TMD remains circularly polarized, indicating that the exciton is both excited and radiatively recombines in the same valley \cite{Xiao2012, Mak2012, Zeng2012,Xu2014, Schaibley2016}. This property makes this class of materials ideal for optical manipulation and has led to the notion of valleytronic devices \cite{Mak2012, Zeng2012, Hsu2015, Hao2016, Yan2016, Wang2016, Neumann2017, Dey2017, Ye2017}.

An inevitability of TMD samples, as in any material, is the presence of defects, which can strongly affect material properties and device performance~\cite{Freysoldt2014, Hong2015, Zhong2016, Khan2017}. Optically, transitions between defect states and bulk states can give rise to new exciton features~\cite{Attacalite2011}. In TMDs, the most abundant point defects are reported to be chalcogen vacancies~\cite{Zhou2013, Zhong2016}, and their presence is believed to modify the TMDs' electronic structure and optical spectrum as evident by low-energy sub-optical gap features in the photoluminescence (PL) spectra~\cite{Tongay2013}; such defects are also thought to degrade valley polarization~\cite{Wang2013,Kim2015}. Recent experiments report that the PL intensity of both the A exciton peak and sub-optical gap features appear to increase with defect density~\cite{Tongay2013, Chow2015} and that excitons localized at defects can behave as single-photon emitters~\cite{Ross2014,Koperski2015,He2015,Chakraborty2015,Palacios-Berraquero2016}.

Much of the current theoretical understanding of defects in TMDs comes from density functional theory (DFT)~\cite{Kohn1965} calculations, which predict that chalcogen vacancies give rise to localized, unoccupied in-gap single-particle states~\cite{Liu2013, Noh2014, Guo2015, Mahjouri-Samani2016, Haldar2015, Komsa2015, Hong2015, Pandey2016, Li2016}, which in turn can affect intervalley scattering probability~\cite{Pulkin2016, Kaasbjerg2017}. A recent calculation~\cite{Naik2017} shows that quasiparticle self-energy corrections arising from many-electron interactions at the GW level do not qualitatively alter this picture. However, to predict changes in the optical spectrum, and to understand how defects alter TMD photophysics, it is necessary to go beyond a quasiparticle (QP) picture and include the electron-hole interactions, which give rise to excitonic effects. 

In this letter, we use the \textit{ab initio} GW plus Bethe-Salpeter equation (GW-BSE) approach~\cite{Hybertsen1986, Rohlfing1998, *Rohlfing2000} as implemented in the BerkeleyGW package~\cite{Deslippe2012} to study the optical spectra, including excitonic effects, of chalcogen vacancy point defects in TMDs. We find that the in-gap defect states give rise to low-energy, exciton states in good agreement with sub-optical gap features seen in PL experiments. Moreover, the similar energy difference between the unoccupied and occupied QP defect states and the QP gap of pristine TMDs results in strong hybridization between the A exciton and defect excitons in the vicinity of the chalcogen vacancy. Remarkably, the A peak excitation energy remains essentially unchanged by the presence of the vacancies, even as the excitons associated with the peak acquire a large degree of defect character. This hybridization leads to significant valley depolarization, suggesting intriguing pathways for controlling optical features and valley polarization through defect engineering, as well as routes to probe defect structure through simple optical measurements.

We start by calculating the QP bandstructure and energy levels of two monolayer TMDs, MoSe$_2$ and WS$_2$, with chalcogen vacancies. We construct a supercell consisting of 5 unit cells along each crystalline-axis of the monolayer plane and then remove a single chalcogen atom (Fig.~\ref{fig:fig1}(a)); we refer to these as 5$\times$5 supercells, and they correspond to a 2\% vacancy concentration. The 5$\times$5 supercell is the smallest supercell that approximates satisfactorily an isolated defect (see SI). We fix the value of the lattice vector to the experimental value at room temperature and then relax the atomic coordinates using DFT within the local density approximation (LDA)~\cite{Kohn1965}. We then use the DFT wavefunctions as a starting point for a one-shot G$_0$W$_0$ calculation~\cite{Hybertsen1986,Deslippe2012}. Dynamical effects in the screening are accounted for within the Hybertsen-Louie generalized plasmon pole (HL-GPP) model~\cite{Hybertsen1986}. We find that inclusion of the full frequency dependence of the dielectric screening and the use of different mean-field starting points does not significantly change our results. Additional computational details can be found in the Supplemental Materials.

The QP bandstructure of MoSe$_2$ with a Se vacancy in the 5$\times$5 supercell BZ is shown in Fig.~\ref{fig:fig1}(b). We find that the chalcogen vacancy results in a single occupied defect state in the valence band (not counting spin degeneracy), which we label as $v_D$, and two nearly degenerate unoccupied defect states in the gap, which we will refer to as $c_{D1}$ and $c_{D2}$, in good agreement with previous calculations~\cite{Naik2017}. Isosurfaces of the square of the wavefunction of these states at $\bar{K}$ are also plotted in Fig.~\ref{fig:fig1}(b).  The defect QP states are localized around the defect site, and their character consists primarily of transition metal $d$-orbitals.  

An energy level diagram of the defect states and the bulk-like valence and conduction states is shown in Fig.~\ref{fig:fig1}(c) for both MoSe$_2$ and WS$_2$. The GW correction opens the gap, but it does not change the qualitative picture of the occupied and unoccupied defect states.  For MoSe$_2$, the QP gap of the bulk states is 2.3 eV, and for WS$_2$ the QP gap of the bulk states is 2.8 eV, which are similar to previous calculations of the QP gap in the defect-free monolayer~\cite{Ugeda2014,Bradley2015,Ashwin2012,Shi2013}. In WS$_2$, spin-orbit coupling splits the in-gap QP states by 0.2~eV. For both MoSe$_2$ and WS$_2$, GW corrections push the occupied and unccupied defect states down in energy with respect to the bulk occupied and unoccupied band edges. The QP energy difference between the unoccupied and occupied defect states is 2.1 eV in MoSe$_2$ and 2.2 eV in WS$_2$. The energy gap between the defect states is quite close to the energy gap between the bulk states, suggesting that the electron-hole interaction can mix defect and bulk transitions significantly, as verified by our GW-BSE calculations below.

\begin{figure}
\begin{center}
\includegraphics[width=0.5\textwidth]{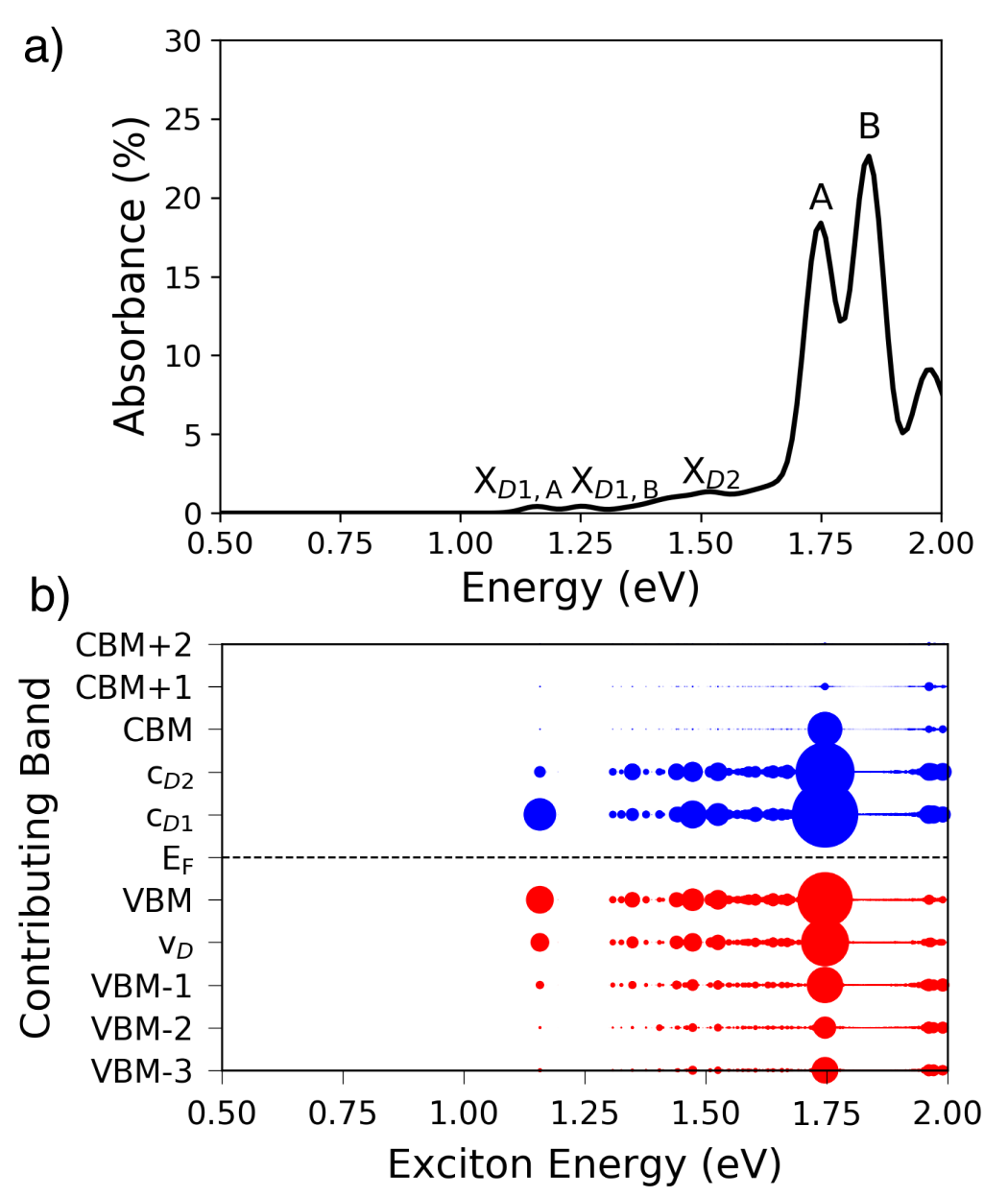}
\centering
\caption{a) Absorbance spectrum of 5x5 supercells of MoSe$_2$ with a single chalcogen vacancy. b) Contributions of each band to each exciton state, plotted with respect to the exciton excitation energy (Only the A-series of spin-orbit split excitons is included for clarity). The bands are labelled as either bulk-like dispersive bands relative to the valence band maximum (VBM) and conduction band minimum (CBM) or as flat defect bands $v_{D}$, $c_{D1}$, and $c_{D2}$, with the occupied band contributions in red and unoccupied band contributions in blue. The size of each dot is proportional to square of the k-space electron-hole amplitude of the contribution from each band weighted by the oscillator strength of the exciton state.}
\label{fig:fig2}
\end{center}
\end{figure}

Next, we examine the excitons of a system with a chalcogen vacancy in the 5$\times$5 supercell of MoSe$_2$ by solving the BSE~\cite{Rohlfing1998, *Rohlfing2000}. Reciprocal space is sampled with the clustered sampling interpolation (CSI) method~\cite{Jornada2017} for an effective k grid of 18$\times$18$\times$1. These parameters allow us to converge linear absorption spectra covering the range of 0 to 2.0~eV and the exciton excitation energies to within 0.15~eV. Spin orbit coupling is included as a perturbation following Refs.~\cite{Qiu2013,Qiu2016}. Additional computation details are provided in the Supplemental Materials.

Fig.~\ref{fig:fig2}(a) shows our calculated absorbance spectrum, namely the amount of light absorbed by the examined monolayer MoSe$_2$ with a single selenium vacancy in the 5$\times$5 supercell. The well-known spin-orbit split A and B peaks are labeled accordingly. The defect introduces additional low-energy features, which we label $X_{D1,A}$, $X_{D1,B}$, and $X_{D2}$. To elucidate these features we break down each exciton state into its component transitions. That is, the exciton wavefunction can be written as a linear combination of electron-hole pairs
\begin{equation}
\Psi^{S}(\mathbf{r}_e,\mathbf{r}_h)=\sum_{vc\mathbf{k}} A^{S}_{vc\mathbf{k}} \psi_{c\mathbf{k}}(\mathbf{r}_e)\psi^*_{v\mathbf{k}}(\mathbf{r}_h),
\end{equation}
where $\mathbf{r}_e$($\mathbf{r}_h$) is the position of the electron(hole); S indexes the exciton state; $v$ and $c$ index respectively the occupied and unoccupied bands; $\psi_{c\mathbf{k}}$ is the wavefunction of the electron in state $c\mathbf{k}$; $\psi_{v\mathbf{k}}$ is the wavefunction of the electron missing from state $v\mathbf{k}$; and  $A^{S}_{vc\mathbf{k}}$ is the electron-hole amplitude. 

Fig.~\ref{fig:fig2}(b) shows the contribution of each occupied band and unoccupied band to a given exciton state. Each exciton is represented by a column of dots, and the size of each dot is proportional to the square of the electron hole amplitude of the contribution from each band weighted by the oscillator strength. The size of each dot goes as (oscillator strength)$\times\sum_{c\mathbf{k}} |A^{S}_{vc\mathbf{k}}|^2$ for each of the occupied bands and (oscillator strength)$\times\sum_{v\mathbf{k}} |A^{S}_{vc\mathbf{k}}|^2$ for each of the unoccupied bands (see SI for further explanation). From this plot, we see that the lowest energy feature, $X_{D1,A}$, at 1.2~eV comes primarily from transitions between the VBM and the unoccupied defect band $c_{D1}$, which make up 90\% of the band to band transitions composing the exciton. $X_{D1,B}$ is the spin-orbit split counterpart of $X_{D1,A}$ and has the same character (not shown in Fig.~\ref{fig:fig2}(b)). The binding energy of the $X_{D1}$ exciton is 0.6 eV, similar to the binding energy of the $A$ exciton in the pristine monolayer, but the radius of the $X_{D1}$ exciton is about 0.6 nm, which is roughly half the size of the exciton in the pristine monolayer~\cite{Qiu2013,Ugeda2014}. The third low-energy feature at 1.5 eV in Fig.~\ref{fig:fig2}(a), $X_{D2}$, includes additional weight from transitions from the occupied defect states to the unoccupied defect states. Unlike $X_{D1}$, which is mainly localized in the $K$ and $K'$ valleys, $X_{D2}$ is highly delocalized in the Brillouin zone (see SI), suggesting that it has a defect-like character. Finally, although peaks A and B occur at roughly the same energy as A and B in the pristine monolayer, they both mix significantly with transitions involving the defect states. This mixing is a consequence of the small difference (on the order of the exciton binding energy) between the energies of the occupied to unoccupied defect transition and the pristine QP gap. We note that the charged defect also has a similar defect-defect energy gap and should thus be expected to participate in similar hybridization with the pristine excitons (see SI). Hybridization with the defect also results in reduction of the energy separation between the A and B peaks (resulting from spin-orbit coupling), since the spin-orbit splitting of the defect bands is considerably smaller than that of the VBM of the pristine monolayer.

\begin{figure}
\begin{center}
\includegraphics[width=0.45\textwidth]{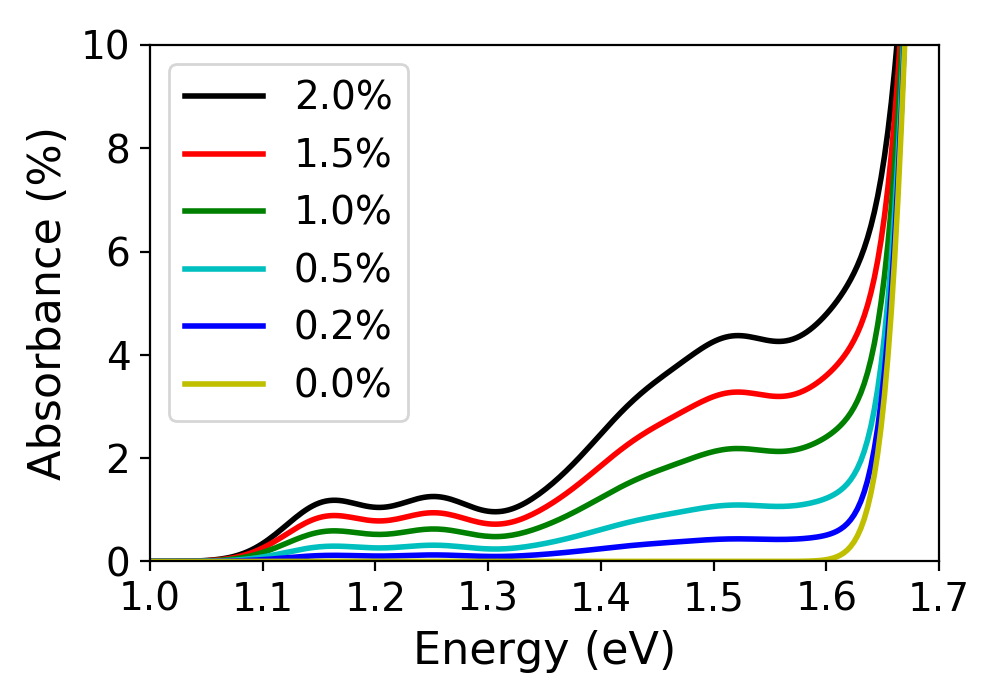}
\centering
\caption{Absorbance spectra of MoSe$_2$ computed for different defect densities. 2.0\% defect density corresponds to the explicitly calculated 5$\times$5  supercell (black line) and 0.0\% corresponds to the pristine monolayer (yellow line). Other densities are evaluated following Eq. 5 in the SI. }
\label{fig:fig3}
\end{center}
\end{figure}

To better understand the effect of defect density on the optical spectrum, we take advantage of the superposition property for linear response to calculate the absorbance at different defect densities (see SI). The resulting absorbance for different defect densities is shown in Fig. \ref{fig:fig3}. As expected, the absorbance of low-lying defect excitons increases significantly with defect density, suggesting a possible optical marker for identification of defect concentration at different regions of the sample.

\begin{figure}
\begin{center}
\includegraphics[width=0.45\textwidth]{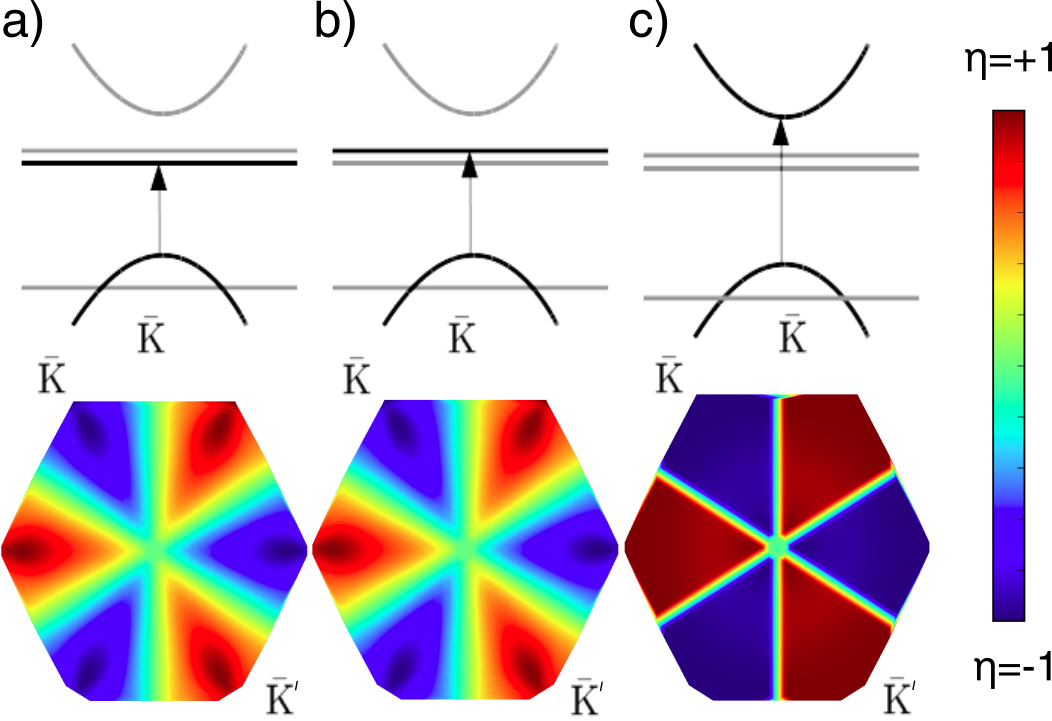}
\centering
\caption{Schematic of band-to-band transition near $\bar{K}$ (top panel) and degree of circular dichroism in the supercell BZ (bottom panel) for the band-to-band transitions a) VBM to $c_{D1}$, b)VBM to $c_{D2}$, and c) VBM to CBM of the monolayer MoSe$_2$ 5$\times$5 supercell with the chalcogen vacancy. The selected band-to-band transition is in black with other bands in gray.}
\label{fig:fig4}
\end{center}
\end{figure}

Finally, we consider how the defect might affect the valley selection rules for left and right circularly polarized light for the A and B excitons. First, we calculate the degree of circular dichroism ($\eta_{vc}(\mathbf{k})$)for the band-to-band transitions for our 5$\times$5 supercells with the chalcogen vacancy (see SI). Fig.~\ref{fig:fig4} shows $\eta_{vc}(\mathbf{k})$ for several band-to-band ($vc$) transitions (for the A series bands) in the BZ of the supercell. Each contour plot of $\eta_{vc}(\mathbf{k})$ is accompanied by a schematic of the bandstructure near $\bar{K}$ with the band-to-band transition denoted by an arrow. We see that transitions between the VBM and the unoccupied defect bands (Fig.~\ref{fig:fig4}(a-b)) have the same optical selection rules as the bulk VBM to CBM transition (Fig.~\ref{fig:fig4}(c)) in the vicinity of the $\bar{K}$ and $\bar{K}'$ points with a somewhat reduced value of $\eta(\mathbf{k})$. On the other hand, since defect states lack the intrinsic crystal symmetry  required to give rise to the coupling of valley and optical helicity  (in fact, they are k-independent in the dilute limit), the transitions between the occupied defect band and the two unoccupied defect bands cannot exhibit any valley-selective circular dichroism.

\begin{figure}
\begin{center}
\includegraphics[width=0.5\textwidth]{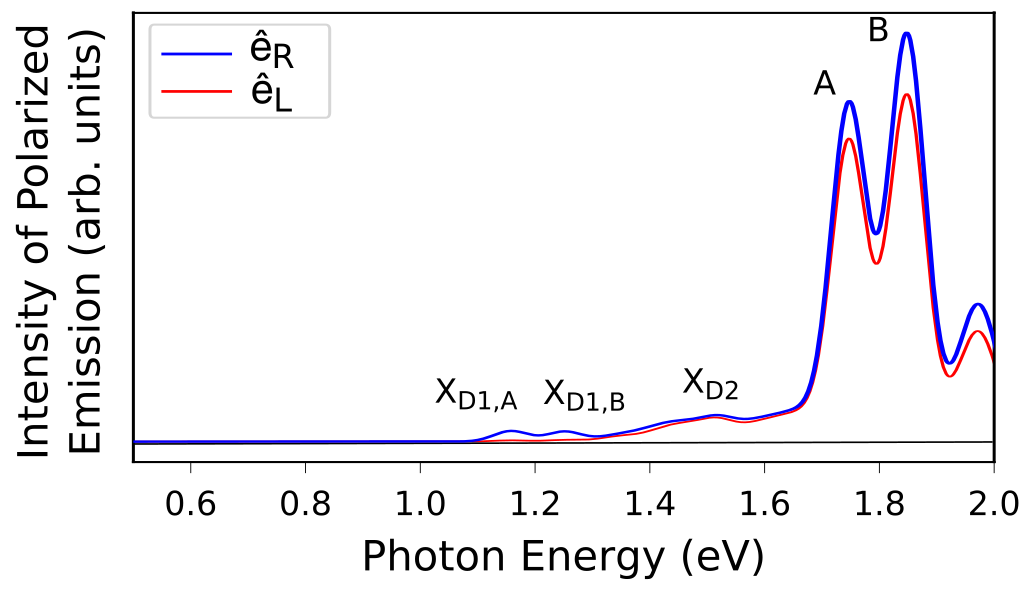}
\centering
\caption{Intensity (arbitrary units) of the instantaneous emission of left-hand circularly polarized light ($\sigma_L$, red) or right-hand circularly polarized light ($\sigma_R$, blue) from an exciton state excited by right-hand circularly polarized light, in a 5x5 unit cell of MoSe$_2$ with a single chalcogen vacancy.}
\label{fig:fig5}
\end{center}
\end{figure}

Next, we include excitonic effects in our analysis of the sensitivity of the valley polarization to defects. Fig.~\ref{fig:fig5} shows the probability that right-hand or left-hand circularly polarized light will be emitted from an exciton excited by right-hand circularly polarized light. We see that the $X_{D1,A}$ and $X_{D1,B}$ features exhibit a high degree of difference in their emission of right and left circularly polarized light, as they arise primarily from VBM to defect state transitions that exhibit a high degree of circular polarization in our analysis of the non-interacting transitions (Fig.~\ref{fig:fig4}(a-b)). A similar circular dichroism from defect states has been seen experimentally for W vacancies in WS$_2$.~\cite{Zhang2017}  $X_{D2}$, on the other hand, exhibits a very small amount of circular dichroism, since it arises primarily from transitions between defect states. Similarly, significant hybridization with defect states results in a dramatic decrease in the expected valley polarization of the A and B excitons, so that the probability of emitting left-polarized light is only slightly less than the probability of emitting right-polarized light. The difference between the two curves for the B peak is slightly larger than for the A peak, since the B exciton hybridizes slightly less with the defect. This suggests that hybridization with defect states could act as a significant source of valley depolarization in TMDs. This mechanism allows for valley depolarization without phonon-assisted intervalley scattering, suggesting that the valley depolarization will be present even at low temperatures, and may explain the plateau in the degree of valley polarization with temperature, which is observed at temperatures below 90 K~\cite{Zeng2012}. We emphasize that this hybridization is a purely excitonic effect that cannot be rationalized within a noninteracting inter-band transitions picture. We note that hybridization as a mechanism for loss of valley polarization might be suppressed by passivating defects in a way that changes defect transition energy levels so that they are either much higher or much lower than the QP gap. It may be further possible to engineer the degree of valley selectivity by coupling to defects with chiral optical selection rules.

In summary, we have performed first-principles GW-BSE calculations of the QP bandstructure and optical spectra of TMD systems with chalcogen vacancies at low concentrations. We find that valence band-to-defect state and defect state-to-defect state transitions give rise to low-energy excitonic features in the optical spectrum in good agreement with the energy of defect-assigned features in PL experiments. Moreover, we find that the similar energy of the gap between the occupied and unoccupied defect states and the QP gap of the pristine system gives rise to strong hybridization between excitons of the pristine system and the defect states. This hybridization dramatically reduces the valley-selective circular polarization of the A and B excitons. The predictive nature of these results can be generalized to other monolayer TMDs, where defects introduce additional energy levels (from localized states) whose separation between occupied and unoccupied levels  is on the scale of the QP gap, suggesting intriguing new pathways for controlling optical features and valley polarization through defect engineering, as well as ways to probe complicated defect structure through optical measurements. 

We thank Felipe H. da Jornada, Jack Deslippe and Mauro Del Ben for valuable discussions. This work was supported by the Center for Computational Study of Excited State Phenomena in Energy Materials, which is funded by the U.S. Department of Energy, Office of Science, Basic Energy Sciences, Materials Sciences and Engineering Division under Contract No. DE-AC02-05CH11231, as part of the Computational Materials Sciences Program. Work performed at the Molecular Foundry was also supported by the Office of Science, Office of Basic Energy Sciences, of the U.S. Department of Energy under the same contract number. S.R.A acknowledges Rothschild and Fulbright fellowships.  This research used resources of the National Energy Research Scientific Computing Center (NERSC), a DOE Office of Science User Facility supported by the Office of Science of the U.S. Department of Energy under Contract No. DE-AC02-05CH11231.

\bibliography{DefMoSe2.bib}

\end{document}